\documentclass [a4paper,fleqn,12pt]{article}
\usepackage{graphicx}
\usepackage {subfigure,epsfig}

\usepackage {amsmath} \usepackage{amssymb} \usepackage{cite} \usepackage{amsthm}
\numberwithin{equation}{section} \numberwithin{table}{section} \mathindent=0pt
\theoremstyle{plain} \newtheorem{theorem}{Theorem}

\numberwithin{theorem}{section}

\begin{document}

\title{\textbf{Special polynomials associated with the $K_2$ hierarchy}}

\author{Nikolai A. Kudryashov}

\date{Department of Applied Mathematics\\
Moscow Engineering and Physics Institute\\ (State University)\\
31 Kashirskoe Shosse, 115409, Moscow, \\ Russian Federation}
\maketitle

\begin{abstract}

New special polynomials associated with the rational solutions of
analogue to the Painlev\'e hierarchies are introduced. The Hirota
relations for these special polynomials are found. Differential -
difference hierarchies for finding special polynomials are
presented. These formulae allow us to search the special polynomials
associated with the hierarchy studied. The ordinary differential
hierarchy for the Yablonskii - Vorob'ev polynomials is given.

\end{abstract}



\section{Introduction}

In this paper our interest is in special polynomials associated with
rational solutions of the hierarchy \cite{Kudryashov01,
Kudryashov02, Kudryashov03}
\begin{equation}\begin{gathered}
\label{ku0.1}
\left(\frac{d}{dz}+w\right)H_N\left[w_z-\frac12\,w^2\right]-z\,w-\beta=0
\end{gathered}
\end{equation}
Here the operator $H_{N}$ is determined by the following recursion
formula \cite{Weiss01, Kudryashov01, Kudryashov02, Kudryashov03}
\begin{equation}\begin{gathered}
\label{ku0.2}H_{N+2}=J[v]\,\Omega[v]\,H_{N}
\end{gathered}\end{equation}
under the conditions
\begin{equation}\begin{gathered}
\label{ku0.3}H_{0}[v]=1,\,\,\quad\,\,H_{1}[v]=v_{zz}+4\,v^2,
\end{gathered}\end{equation}
where the operatops $\Omega[v]$ and $J[v]$ take the form
\cite{Weiss01, Kudryashov01, Kudryashov02, Kudryashov03}
\begin{equation}\begin{gathered}
\label{ku0.4}\Omega=D^3+2\,v\,D+v_z,\,\,\quad\,D=\frac{d}{dz}
\end{gathered}\end{equation}
\begin{equation}\begin{gathered}
\label{ku0.5}J=D^3+3\left(vD+D\,v\right)+2\,\left(D^2\,v\,D^{-1}+D^{-1}\,v\,D^2\right)+
\\
+8 \left(v^2\,D^{-1}+D^{-1}\,v^2\right),\,\quad\, D^{-1}=\int\,dz
\end{gathered}\end{equation}

Hierarchy \eqref{ku0.1} is important because the special solutions
of the Fordy - Gibbons equation \cite{Fordy01}, the
Caudrey-Dodd-Gibbon hierarchy (Savada - Kotera
hierarchy)\cite{Caudrey01, Savada01} and the Kaup - Kupershmidt
hierarchy \cite{Kaup01} can be expressed through solutions of the
hierarchy \eqref{ku0.1}. Hierarchy \eqref{ku0.1} is called the $K_2$
hierarchy in \cite {Kudryashov02} taking into consideration jubilee
year in 2000 by Kovalevskaya and Kruskal. Apparently hierarchy
\eqref{ku0.1} defines new transcendental functions like the
Painlev\'e equations do.

Hierarchy \eqref{ku0.1} can be written in the form as well
\begin{equation}\begin{gathered}
\label{ku0.6}-\frac12
\left(\frac{d}{dz}-2\,w\right)G_N\left[-2\,w_z-2\,w^2\right]-z\,w-\beta=0
\end{gathered}
\end{equation}
The recursion relations  $G_{N}$ is determined by the
operator\cite{Kudryashov01,Kudryashov02,Kudryashov03, Weiss01}
\begin{equation}\begin{gathered}
\label{ku0.7}G_{N+2}=J_{1}[u]\,\Omega[u]\,G_{N}
\end{gathered}\end{equation}
under the conditions
\begin{equation}\begin{gathered}
\label{ku0.8}G_{0}[u]=1,\,\,\quad\,G_{1}[u]=u_{zz}+\frac{1}{4}\,u^2
\end{gathered}\end{equation}
The operator $J_{1}[u]$ takes the form
\begin{equation}\begin{gathered}
\label{ku0.9}J_{1}=D^3+\frac{1}{2}\left(D^2\,u\,D^{-1}+D^{-1}\,u\,D^2\right)+
\\
+ \frac{1}{8} \left(u^2\,D^{-1}+D^{-1}\,u^2\right)
\end{gathered}\end{equation}

Hierarchy \eqref{ku0.1} is similar in appearance to the second
Painlev\,e hierarchy but hierarchy \eqref{ku0.1} is distinguished
from the $P_{2}$ hierarchy by the operator $H_{N}$ in \eqref{ku0.1}
\cite{Kudryashov04, Clarkson01, Demina01}. Special polynomials
associated with the rational solutions of the $P_{2}$ hierarchy were
considered in papers \cite{Clarkson01, Demina01}. Recently it was
shown \cite{Kudryashov05} that the rational solutions of hierarchy
\eqref{ku0.1} at $N=1$ can be found using the special polynomials
$Q^{(1)}_{n}(z)$ and $R^{(1)}_{n}(z)$ by the formulae
\begin{equation}\begin{gathered}
\label{ku0.5a}w(z;\beta^{(1)}_{n})=(-1)^{n}\frac{d}{dz}\,\ln{\frac{Q^{(1)}_{n-1}}
{Q^{(1)}_{n}}},\,\,\,\quad\,w(z;\beta^{(2)}_{n})=(-1)^{n-1}\frac{d}{dz}\,
\ln{\frac{R^{(1)}_{n-1}} {R^{(1)}_{n}}}
\end{gathered}
\end{equation}
These special polynomials $Q^{(1)}_{n}(z)$ and $R^{(1)}_{n}(z)$ were
obtained in \cite{Kudryashov05} taking into consideration the power
expansions near infinity \cite{Demina01, Demina02, Kudryashov06}.
This approach is not simple for calculating the special polynomials.
This raises the question of whether the recursion formulae for
finding the special polynomials $Q^{(1)}_{n}(z)$ and
$R^{(1)}_{n}(z)$.

The aim of this paper is to introduce the special polynomials
$Q^{(N)}_{n}(z)$ and $R^{(N)}_{n}(z)$ associated with the rational
solutions of hierarchy \eqref{ku0.1} and to derive the recursion
formulae for finding the special polynomials $Q^{(N)}_{n}(z)$ and
$R^{(N)}_{n}(z)$.

The main result of this paper is the differential - difference
hierarchies
\begin{equation}\begin{gathered}
\label{ku0.5b}P_{n+1}\,P_{n-1}=P_{n}\,F_{N}\left[A\frac{d^2}{dz^2}\,\ln{P_{n}}\right]
\end{gathered}
\end{equation}
for the polynomials $P_{n}(z)\equiv Q^{(N)}_{n}(z)$ (and
$P_{n}(z)\equiv R^{(N)}_{n}(z)$). Here $A$ is a constant, $F_{N}$ is
a operator which depends on the operators $H_n[v]$ or $G_n[u]$. This
formula allows us to look for special polynomials associated with
rational solutions of hierarchy \eqref{ku0.1} using
$P_{0}(z)=Q^{(N)}_{0}(z)=R^{(N)}_{0}(z)=1$, $P_{1}(z)=Q^{(N)}_{1}=z$
and $P_{1}(z)=R^{(N)}_{1}(z)=z^2$.

This paper is organized as follows. The general properties of the
hierarchy considered are presented in section 2. Some relations for
the special polynomials $Q^{(N)}_{n}(z)$ and $R^{(N)}_{n}(z)$ and
the differential - difference hierarchies for finding polynomials
$Q^{(N)}_{n}(z)$ and $R^{(N)}_{n}(z)$ are given in section 3. The
special polynomials $Q^{(N)}_{n}(z)$ and $R^{(N)}_{n}(z)$ associated
with the rational solutions of the first, the second and the third
members of hierarchy \eqref{ku0.1} are introduced in sections 4, 5
and 6.

\section{Some properties of hierarchy \eqref{ku0.1}}

Let us briefly review some facts concerning equation \eqref{ku0.1}
needed later. Suppose $w(z)\equiv w(z;\beta)$ is a solution of
\eqref{ku0.1}.

Assuming
\begin{equation}\begin{gathered}
\label{ku0.16}w(z)=\frac{\varphi_{zz}}{\varphi_{z}}
\end{gathered}\end{equation}
and taking into account
\begin{equation}\begin{gathered}
\label{ku0.17}
w_z-\frac{1}{2}\,w^2=\{\varphi;z\}=\frac{\varphi_{zzz}}{\varphi_{zz}}-
\frac{3\,\varphi^{2}_{zz}}{2\,\varphi^{2}_{z}}
\end{gathered}\end{equation}
we have the equation in the form \cite{Kudryashov02,Kudryashov03}
\begin{equation}\begin{gathered}
\label{ku0.18}H_{N}\left[\{\varphi;z\}\right]-z-\left(\beta-1\right)
\frac{\varphi}{\varphi_z}=0
\end{gathered}\end{equation}
Using
\begin{equation}\begin{gathered}
\label{ku0.19}w(z)=-\frac{\psi_{zz}}{2\psi_{z}}
\end{gathered}\end{equation}
and taking into consideration
\begin{equation}\begin{gathered}
\label{ku0.20}
-2\,w_z-2\,w^2=\{\psi;z\}=\frac{\psi_{zzz}}{\psi_{zz}}-
\frac{3\,\psi^{2}_{zz}}{2\,\psi^{2}_{z}}
\end{gathered}\end{equation}
we obtain the equation in the form
\begin{equation}\begin{gathered}
\label{ku0.21}G_{N}\left[\{\psi;z\}\right]-z-\left(\beta-1\right)
\frac{\psi}{\psi_z}=0
\end{gathered}\end{equation}

The Backlund transformations for solutions of hierarchy
\eqref{ku0.1} can be written in the form \cite{Kudryashov02,
Kudryashov03}
\begin{equation}\begin{gathered}
\label{ku0.22}w(z,2-\beta)=w(z,\beta)-\frac{2\,\beta-2}{H_{N}
\left[w_z-\frac12\,w^2\right]-z}
\end{gathered}\end{equation}
and
\begin{equation}\begin{gathered}
\label{ku0.23}w(z,-1-\beta)=w(z,\beta)-\frac{2\,\beta+1}{G_{N}
\left[-2\,w_z-2\,w^2\right]-z}
\end{gathered}\end{equation}
These transformations allow us to find the rational solutions of
hierarchy \eqref{ku0.1}\cite{Kudryashov02, Kudryashov03}. These
formulae can be presented in the form
\begin{equation}\begin{gathered}
\label{ku0.25}{G_{N}
\left[-2\,w_z-2\,w^2\right]-z}=-\frac{2\beta+1}{y(z;-1-\beta)-y(z;\beta)}
\end{gathered}\end{equation}
and
\begin{equation}\begin{gathered}
\label{ku0.24}{H_{N}
\left[w_z-\frac12\,w^2\right]-z}=-\frac{2\,(\beta-1)}{y(z;2-\beta)-y(z;\beta)}
\end{gathered}\end{equation}

The rational solutions of hierarchy \eqref{ku0.1} are classified in
the following theorem.
\begin{theorem}
\label{T:2.1} Equations \eqref{ku0.1} possesses rational solutions
if and only if $\beta \in \mathbb{Z}/\{1\pm3k$, $k \in
\mathbb{N}\cup0\}$. They are unique and have the form
\begin{equation}
\begin{gathered}
\label{2.4}w(z;\beta_n^{(1)})=(-1)^{n}\frac{d}{dz}\ln
{\left(\frac{Q^{(N)}_{n-1}}{Q^{(N)}_n} \right)},\\
w(z;\beta_n^{(2)})=(-1)^{n-1}\frac{d}{dz}\ln
{\left(\frac{R^{(N)}_{n-1}} {R^{(N)}_n}\right)},
\end{gathered}
\end{equation}
where $Q^{(N)}_n(z)$ and $R^{(N)}_n(z)$ are polynomials, $n\in
\mathbb{N}$ and
\begin{equation}
\begin{gathered}
\label{2.5}\beta_{n}^{(1)}=(-1)^n\,\left(3\left[\frac{n+1}{2}\right]-
1+(-1)^n\right),\\
\beta_{n}^{(2)}=(-1)^{n+1}\,\left(3\left[\frac{n}{2}\right]+1+(-1)^
{n+1}\right)
\end{gathered}
\end{equation}
with $[x]$ denoting the integer part of $x$. The only remaining
rational solution is the trivial solution $w(z;0)=0$.
\end{theorem}
\begin{proof}This theorem can be proved by the analogy with corresponding theorem
of the work \cite{Kudryashov05}. The rational solutions of
\eqref{ku0.1} can be described with the help of two families of
polynomials. The polynomials $\{Q^{(N)}_n(z)\}$ we call the first
family and $\{R^{(N)}_n(z)\}$ the second. By $p_n^{(1)}$
$(p_n^{(2)})$ denote the degree of $Q^{(N)}_n(z)$ $(R^{(N)}_n(z))$.
The special polynomials $Q^{(N)}_n(z)$ and $R^{(N)}_n(z)$ can be
defined as monic polynomials and each polynomial can be presented in
the form
\begin{equation}\begin{gathered}
\label{2.9}Q^{(N)}_n(z)=\sum_{k=0}^{p_{n}^{(1)}}A^{(1,N)}_{n,k}\,z^{p_{n}^{(1)}\,-\,k},\qquad
A^{(1,N)}_{n,0}=1,\\
R^{(N)}_n(z)=\sum_{k=0}^{p_{n}^{(2)}}A^{(2,N)}_{n,k}\,z^{p_{n}^{(2)}\,-\,k},\qquad
A^{(2,N)}_{n,0}=1.
\end{gathered}
\end{equation}
The first non-trivial solutions of \eqref{ku0.1} are $w(z;-1)=1/z$
and $w(z;2)=-2/z$. Hence it can be set
$Q^{(N)}_0(z)=R^{(N)}_0(z)=1$, $Q^{(N)}_1(z)=z$, $R^{(N)}_1(z)=z^2$.

We also observe that degree of polynomials can be rewritten in terms
of $\beta_{n}^{(j)}$
\begin{equation}\begin{gathered}
\label{2.21a}p_{n}^{(1)}=\sum_{i=1}^{n}|\beta_i^{(1)}|=\frac{k(k+1)}{2}-\frac12\,
\left[\frac{k+1}{3}\right]-
\frac32\left[\frac{k+1}{3}\right]^2,\,k\stackrel{def}{=} |\beta_n^{(1)}|,\hfill\\
p_{n}^{(2)}=\sum_{i=1}^{n}|\beta_i^{(2)}|=\frac{k(k+1)}{2}+\frac12\,\left[\frac{k+2}
{3}\right]- \frac32\left[\frac{k+2}{3}\right]^2,\,k\stackrel{def}{=}
|\beta_n^{(2)}|.
\end{gathered}
\end{equation}
\end{proof}
The first few $\beta_{n}^{(j)}$ and $p_n^{(j)}$ are given in Table
\ref{t:1}.

\begin{table}[h]%[t h]
    \center
    \caption{Values of $\beta_{n}^{(j)}$ and $p_n^{(j)}$ $(j=1,2)$. } \label{t:1}
    \begin{tabular}{|c|c|c|c|c|c|c|c|c|c|c|c|c|} %{||c|c|c|c|c|c|c||}     %{||c|c|p{65mm}||}
        \hline
        $ n $& $1$ &$2$ &$3$ & $4$ & $5$ & $6$ & $7$ & $8$ & $9$ & $10$ & $11$ &
        $12$\\ \hline
        $ \beta_{n}^{(1)} $ & $-1$ &$3$ & $-4$ & $6$ & $-7$ & $9$ & $-10$ & $12$ & $-13$
        & $15$ & $-16$ &
        $18$\\ \hline
        $ p_{n}^{(1)} $ & $1$ &$4$ & $8$ & $14$ & $21$ & $30$ & $40$ & $52$ & $65$
        & $80$ & $96$ &
        $114$\\ \hline
         $ \beta_{n}^{(2)} $ & $2$ &$-3$ & $5$ & $-6$ & $8$ & $-9$ & $11$ & $-12$ &
         $14$ & $-15$ & $17$ &
        $-18$\\ \hline
        $ p_{n}^{(2)} $ & $2$ &$5$ & $10$ & $16$ & $24$ & $33$ & $44$ & $56$ & $70$ &
         $85$ & $102$ & $120$\\
        \hline
    \end{tabular}
\end{table}

\section{Hirota relations and differential - difference hierarchies for special polynomials}

In this section we derive the Hirota relations and the differential
- difference hierarchies for special polynomials $Q^{(N)}_{n}(z)$
and $R^{(N)}_{n}(z)$. Later in this section when it does cause any
contradiction the index $N$ will be omitted.

\begin{theorem}
\label{T:2.1} Special polynomials $Q_n(z)$ and $R_n(z)$ satisfy the
following Hirota relations
\begin{equation}\begin{gathered}
\label{E1.1}D_{z}Q_{2m+2}\bullet
Q_{2m}=\left(4+6\,m\right)\,Q^{\frac12}_{2m}\,
Q_{2m+1}\,Q^{\frac12}_{2m+2},\,\quad\,(m=0,1,...)
\end{gathered}\end{equation}
\begin{equation}\begin{gathered}
\label{E1.2}D_{z}Q_{2m+1}\bullet Q_{2m-1}=(1+6\,m)\,Q^{2}_{2m},\,\,
\quad\,(m=1,2,...)
\end{gathered}\end{equation}
\begin{equation}\begin{gathered}
\label{E1.3}D_{z}R_{2m+2}\bullet R_{2m}=\left(5+6\,m\right)\,
R_{2m+1}^2,\,\quad\,(m=0,1,...)
\end{gathered}\end{equation}
\begin{equation}\begin{gathered}
\label{E1.4}D_{z}R_{2m+1}\bullet
R_{2m-1}=\left(2+6\,m\right)\,R^{\frac12}_{2m-1}\,
R_{2m}\,R^{\frac12}_{2m+1},\,\quad\,(m=1,2,...)
\end{gathered}\end{equation}
where $D_{z}$ is the Hirota operator defined by
\begin{equation}\begin{gathered}
\label{E1.5}D_{z}f(z)\bullet\,g(z)=\left[\left(\frac{d}{dz_1}-\frac{d}{dz_2}
\right)f(z_1)\,g(z_2)\right]_{z_1=z_2=z}
\end{gathered}\end{equation}
\end{theorem}
\begin{proof} Without loss of generality let us prove formulae \eqref{E1.1}
and \eqref{E1.2}. Comparison of
equations \eqref{ku0.18} and \eqref{ku0.25},  \eqref{ku0.21} and
\eqref{ku0.24} yields
\begin{equation}\begin{gathered}
\label{4.3}\varphi=\left(\frac{Q_{2m+2}}{Q_{2m}}\right)^{\frac{1}{2}},\,\,
\quad\,\psi=\frac{Q_{2m+1}}{Q_{2m-1}}
\end{gathered}\end{equation}
Solutions of equations \eqref{ku0.18} and \eqref{ku0.24} can be
found taking into account the following recursion formulae
\cite{Kudryashov02, Kudryashov03, Weiss01}
\begin{equation}\begin{gathered}
\label{4.5}\psi_z=\frac{\varphi^{4}}{\varphi_{z}^2},\,\,\quad\,\,\varphi_{z}=
\frac{\psi}{{\psi_{z}}^{\frac{1}{2}}}
\end{gathered}\end{equation}
Substituting \eqref{4.3} into the second equation \eqref{4.5} yields
\begin{equation}\begin{gathered}
\label{4.8}\left(Q_{2m+2,z}\,Q_{2m}-Q_{2m+2}\,Q_{2m}\right)^2\,
\left(Q_{2m+1,z}\,Q_{2m-1}-Q_{2m+1}\,Q_{2m-1}\right)=\\
=4\,Q_{2m}^3\,Q_{2m+1}^{2}\,Q_{2m+2}
\end{gathered}\end{equation}
Assuming
\begin{equation}\begin{gathered}
\label{4.7}\varphi=\left(\frac{Q_{2m}}{Q_{2m-2}}\right)^
{\frac{1}{2}},\,\,\quad\,\psi=\frac{Q_{2m+1}}{Q_{2m-1}}
\end{gathered}\end{equation}
in the first equation \eqref{4.5} we get
\begin{equation}\begin{gathered}
\label{4.10}\left(Q_{2m+1,z}\,Q_{2m+1}-Q_{2m+1}\,Q_{2m-1}\right)\,
\left(Q_{2m,z}\,Q_{2m-2}-Q_{2m}\,Q_{2m-2}\right)^{2}=\\
=4\,Q_{2m-2}\,Q_{2m-1}^{2}\,Q_{2m}^{3}
\end{gathered}\end{equation}
From \eqref{4.8} and \eqref{4.10} we have
\begin{equation}\begin{gathered}
\label{4.11}\frac{\left(Q_{{2m+2,z}}\,Q_{{2m}}-Q_{{2m+2}}\,Q_{{2m,z}}\right)^2}{Q_{2m}
\,Q^2_{2m+1}\,Q_{2m+2}}=\frac{\left(Q_{{2m,z}}\,Q_{{2m-2}}-Q_{{2m}}\,Q_{{2m-2,z}}
\right)^2}{Q_{2m-2} \,Q^2_{2m-1}\,Q_{2m}}
,\\
\,\quad\,(m=1,2,...)
\end{gathered}\end{equation}
By the induction taking into account $m=1$ and
$Q_2(z)=\left(z^2+C_{1}\right)^2$ (where $C_1$ is constant) in the
right hand side \eqref{4.11} we get
\begin{equation}\begin{gathered}
\label{4.12}{\left(Q_{{2,z}}\,Q_{{0}}-Q_{{2}}\,Q_{{0,z}}
\right)^2}=16\,{Q_{0} \,Q^2_{1}\,Q_{2}}
\end{gathered}\end{equation}
Suppose we have at $m=k$
\begin{equation}\begin{gathered}
\label{4.14}{\left(Q_{{2k,z}}\,Q_{{2k-2}}-Q_{{2k}}\,Q_{{2k-2,z}}\right)^2}=F_{1}(k)\,{Q_{2k-2}
\,Q^2_{2k-1}\,Q_{2k}}
\end{gathered}\end{equation}
then from \eqref{4.11} we obtain the equality at $m=k+1$
\begin{equation}\begin{gathered}
\label{4.14a}{\left(Q_{{2k+2,z}}\,Q_{{2k}}-Q_{{2k+2}}\,Q_{{2k,z}}\right)^2}=F_{1}(k)\,{Q_{2k}
\,Q^2_{2k+1}\,Q_{2k+2}}
\end{gathered}\end{equation}
Taking  into account equation \eqref{4.14a} at $k=m$ we find from
\eqref{4.10}
\begin{equation}\begin{gathered}
\label{4.15}Q_{{2m+1,z}}\,Q_{{2m-1}}-Q_{{2m+1}}\,Q_{2m-1,z}=F_{2}(m)\,Q_{{2m}}^2
\end{gathered}\end{equation}

We determine constants $F_1(m)$ and $F_2(m)$ taking into account
unique polynomials. Finally the relations for $Q_n(z)$ and for
$R_n(z)$ can be written in the form
\begin{equation}\begin{gathered}
\label{4.16}Q_{{2m+2,z}}\,Q_{{2m}}-Q_{{2m+2}}\,Q_{{2m,z}}=
\left(4+6\,m\right)\,Q^{\frac{1}{2}}_{2m}\,Q_{2m+1}\,Q^{\frac12}_{2m+2},\\
\,\quad\,(m=0,1,...)
\end{gathered}\end{equation}
\begin{equation}\begin{gathered}
\label{4.17}Q_{2m+1,z}\,Q_{2m-1}-Q_{2m+1}\,Q_{2m-1,z}=(1+6\,m)\,Q^{2}_{2m},\,\,
\quad\,(m=1,2,...)
\end{gathered}\end{equation}
\begin{equation}\begin{gathered}
\label{4.18}R_{2m+2,z}\,R_{2m}-R_{2m+2}\,
R_{2m,z}=\left(5+6\,m\right)R^2_{2m+1}, \,\,\quad\,(m=0,1,...)
\end{gathered}\end{equation}
\begin{equation}\begin{gathered}
\label{4.19}R_{2m+1,z}\,R_{2m-1}-R_{2m+1}\,R_{2m-1,z}=\left(2+6\,m\right)
R^{\frac{1}{2}}_{2m-1}\,R_{2m}\,R^{\frac12}_{2m+1}\\
\,\,\quad\,(m=1,2,...)
\end{gathered}\end{equation}
Now if we write equations \eqref{4.16}, \eqref{4.17}, \eqref{4.18}
and \eqref{4.19} using the Hirota operator we get equations
\eqref{E1.1},\eqref{E1.2}, \eqref{E1.3} and \eqref{E1.4}.
\end{proof}

\begin{theorem}
\label{T:2.2} The following relations are valid for the special
polynomials $Q_n(z)$ and $R_n(z)$
\begin{equation}\begin{gathered}
\label{E1.10}\frac{d^2}{dz^2}\,\ln{(Q_{2m})}+\frac{D^2_{z}\,{Q_{2m}\bullet
\,Q_{2m-1}}}{Q_{2m}\,Q_{2m-1}}=0, \,\,\quad\,\,(m=1,2,...)
\end{gathered}\end{equation}
\begin{equation}\begin{gathered}
\label{E1.11}\frac{d^2}{dz^2}\,\ln{(Q_{2m})}+\frac{D^{2}_{z}\,{Q_{2m}\bullet
\,Q_{2m+1}}}{Q_{2m+2}\,Q_{2m+1}}=0, \,\,\quad\,\,(m=0,1,...)
\end{gathered}\end{equation}
\begin{equation}\begin{gathered}
\label{E1.12}\frac{d^{2}}{dz^2}\ln{R_{2m+1}}+\frac{D^{2}_{z}\,R_{2m}\bullet\,R_{2m+1}}{R_{2m}\,
R_{2m+1}}\, \,=0, \,\,\quad\,\,(m=0,1,...)
\end{gathered}\end{equation}
\begin{equation}\begin{gathered}
\label{E1.13}\frac{d^{2}}{dz^2}\ln{R_{2m+1}}+\frac{D^{2}_{z}\,R_{2m+1}\bullet
R_{2m+2}}{R_{2m+1}R_{2m+2}}\,=0,\,\, \quad\,\,(m=0,1,...)
\end{gathered}\end{equation}
where $D_z$ is the Hirota operator defined by \eqref{E1.5}.
\end{theorem}
\begin{proof}
To prove theorem \eqref {T:2.2} we use the relations for solutions
$w(z;\beta)$, $w(z;2-\beta)$ and $w(z;-1-\beta)$ in the form
\cite{Kudryashov03}
\begin{equation}\begin{gathered}
\label{4.20}w_{z}(z;\beta)-\frac{1}{2}\,w^{2}(z;\beta)=w_{z}(z;2-\beta)-
\frac{1}{2}\,w^2(z;2-\beta)
\end{gathered}\end{equation}
\begin{equation}\begin{gathered}
\label{4.21}w_{z}(z;\beta)+\,w^{2}(z;\beta)=w_{z}(z;-1-\beta)+
\,w^2(z;-1-\beta)
\end{gathered}\end{equation}
Assuming
\begin{equation}\begin{gathered}
\label{4.22}w(z;\beta)=-\frac{Q_{2m,z}}{Q_{2m}}+\frac{Q_{2m+1,z}}{Q_{2m+1}},\,\,\quad\,
w(z;2-\beta)=\frac{Q_{2m+1,z}}{Q_{2m+1}}+\frac{Q_{2m+2,z}}{Q_{2m+2}}
\end{gathered}\end{equation}
in equality \eqref{4.20} we have
\begin{equation}\begin{gathered}
\label{4.23}2\,\frac{Q_{2m+2,zz}}{Q_{2m+2}}-\frac{Q^2_{2m+2,z}}{Q^2_{2m+2}}-
2\,\frac{Q_{2m,zz}}{Q_{2m}}+\frac{Q^2_{2m,z}}{Q^2_{2m}}
-2\frac{Q_{2m+1,z}\, Q_{2m+2,z}}{Q_{2m+1}\,Q_{2m+2}}
+\\
+2\frac{Q_{2m,z}\, Q_{2m+1,z}}{Q_{2m}\,Q_{2m+1}}=0
\end{gathered}\end{equation}
From \eqref{4.23} we obtain
\begin{equation}\begin{gathered}
\label{4.24}\frac{d^2}{dz^2}\,\ln{Q_{2m+2}}+\frac{D^2_{z}Q_{2m+1}\bullet\,Q_{2m+2}}
{Q_{2m+1}\,Q_{2m+2}}=\frac{d^2}{dz^2}\,\ln{Q_{2m}}+\frac{D^2_{z}Q_{2m+1}\bullet\,Q_{2m}}
{Q_{2m+1}\,Q_{2m}}
\end{gathered}\end{equation}
Substituting
\begin{equation}\begin{gathered}
\label{4.25}w(z;\beta)=\frac{Q_{2m-1,z}} {Q_{2m-1}}-
\frac{Q_{2m,z}}{Q_{2m}},\,\,\quad\,w(z;-1-\beta)=-\frac{Q_{2m,z}}
{Q_{2m}}-\frac{Q_{2m+1,z}}{Q_{2m+1}}
\end{gathered}\end{equation}
in equality \eqref{4.21} yields
\begin{equation}\begin{gathered}
\label{4.30}\frac{Q_{2m-1,z,z}}{Q_{2m-1}}-
\frac{2\,Q_{2m,z}\,Q_{2m-1,z}}{Q_{2m}\,Q_{2m-1}}=\frac{Q_{2m+1,zz}}{Q_{2m+1}}
-\frac{2\,Q_{2m,z}\,Q_{2m+1,z}}{Q_{2m}\,Q_{2m+1}}
\end{gathered}\end{equation}
Adding $\frac{Q_{2m,zz}}{Q_{2m}}$ to both parts of equality
\eqref{4.30} we get
\begin{equation}\begin{gathered}
\label{4.31}\frac{D^{2}_{z}
Q_{2m-1}\bullet\,Q_{2m}}{Q_{2m-1}\,Q_{2m}}=\frac{D^{2}_{z}
Q_{2m}\bullet\,Q_{2m+1}}{Q_{2m}\,Q_{2m+1}}
\end{gathered}\end{equation}
Adding $\frac{2\,Q_{2m,zz}}{Q_{2m}}-\frac{Q^{2}_{2m,z}}{Q^{2}_{2m}}$
to both parts of equality \eqref{4.30} yields
\begin{equation}\begin{gathered}
\label{4.32}\frac{d^2}{dz^2}\,\ln{Q_{2m}}+\frac{D^2_{z}Q_{2m}\bullet\,Q_{2m-1}}
{Q_{2m}\,Q_{2m-1}}=\frac{d^2}{dz^2}\,\ln{Q_{2m}}+\frac{D^2_{z}Q_{2m}\bullet\,Q_{2m+1}}
{Q_{2m}\,Q_{2m+1}}
\end{gathered}\end{equation}
We have
\begin{equation}\begin{gathered}
\label{4.33}\frac{d^2}{dz^2}\,\ln{Q_{0}}+\frac{D^2_{z}Q_{0}\bullet\,Q_{1}}
{Q_{0}\,Q_{1}}=0
\end{gathered}\end{equation}
By the induction assuming at $m=k$ we have
\begin{equation}\begin{gathered}
\label{4.34}\frac{d^2}{dz^2}\,\ln{Q_{2k}}+\frac{D^2_{z}Q_{2k}\bullet\,Q_{2k+1}}
{Q_{2k}\,Q_{2k+1}}=0
\end{gathered}\end{equation}
 then we obtain from \eqref{4.24}
\begin{equation}\begin{gathered}
\label{4.35}\frac{d^2}{dz^2}\,\ln{Q_{2k+2}}+\frac{D^2_{z}Q_{2k+2}\bullet\,Q_{2k+1}}
{Q_{2k+2}\,Q_{2k+1}}=0
\end{gathered}\end{equation}
From \eqref{4.32} we have
\begin{equation}\begin{gathered}
\label{4.36}\frac{d^2}{dz^2}\,\ln{Q_{2k+2}}+\frac{D^2_{z}Q_{2k+2}\bullet\,Q_{2k+3}}
{Q_{2k+2}\,Q_{2k+3}}=0
\end{gathered}\end{equation}
This completes the proof.
\end{proof}

\begin{theorem}
\label{T:2.3} The following differential - difference hierarchies
are valid for the special polynomials $Q_n(z)$ and $R_n(z)$
\begin{equation}\begin{gathered}
\label{E1.20}Q_{2m+2}\,Q_{2m}=Q^{2}_{2m+1}\,\left(H_{N}\left[\frac{3}{2}\frac{d^2}{dz^2}
\ln(Q_{2m+1})\right]-z\right)^{2},\quad(m=0,1,...)
\end{gathered}\end{equation}
\begin{equation}\begin{gathered}
\label{E1.21}Q_{2m+1}\,Q_{2m-1}=-Q^{2}_{2m}\left(G_{N}\left[6\,\frac{d^2}{dz^2}
\ln(Q_{2m})\right]-z\right),\quad\,(m=1,2,...)
\end{gathered}\end{equation}
\begin{equation}\begin{gathered}
\label{E1.22}R_{2m+2}\,R_{2m}=-R^{2}_{2m+1}\left(G_{N}\left[6\frac{d^2}{dz^2}
\ln(R_{2m+1})\right]-z\right),\quad\,(m=0,1,...)
\end{gathered}\end{equation}
\begin{equation}\begin{gathered}
\label{E1.23}R_{2m+1}\,R_{2m-1}=R^{2}_{2m}\,\left(H_{N}\left[\frac{3}{2}\,\frac{d^2}{dz^2}
\ln(R_{2m})\right]-z\right)^{2},\quad\,(m=1,2,...)
\end{gathered}\end{equation}
\end{theorem}

\begin{proof} Without loss of generality let us obtain the formula \eqref{E1.20}.
From \eqref{E1.10} we have the equality
\begin{equation}\begin{gathered}
\label{4.37}\frac{Q_{2m,z,z}}
{Q_{2m}}-\frac{Q^{2}_{2m,z}}{2\,Q^{2}_{2m}}-
\frac{Q_{2m,z}\,Q_{2m+1,z}}{Q_{2m}\,Q_{2m+1}}=-
\frac{Q_{2m+1,zz}}{2\,Q_{2m+1}}
\end{gathered}\end{equation}
Assuming
\begin{equation}\begin{gathered}
\label{4.38}y(z;\beta_n)=-\frac{d}{dz}\ln{\frac{Q_{2m}}{Q_{2m+1}}}
\end{gathered}\end{equation}
we get
\begin{equation}\begin{gathered}
\label{4.39}y_{z}-\frac{1}{2}\,y^2=-\frac{Q_{2m,z,z}}
{Q_{2m}}+\frac{Q^{2}_{2m,z}}{2\,Q^{2}_{2m}} +
\frac{Q_{2m,z}\,Q_{2m+1,z}}{Q_{2m}\,Q_{2m+1}}+\\+\frac{Q_{2m+1,zz}}{Q_{2m+1}}-
\frac{3\,Q^2_{2m+1,z}}{2\,Q^2_{2m+1}}=\frac{3}{2}\frac{d^2}{dz^2}\ln{Q_{2m+1}}
\end{gathered}\end{equation}
Taking into account \eqref{E1.1} we obtain from equation
\eqref{ku0.24}
\begin{equation}\begin{gathered}
\label{4.40}H_{N}\left[\frac{3}{2}\frac{d^2}{dz^2}\,\ln{Q_{2m+1}}\right]-z=
\frac{Q^{\frac{1}{2}}_{2m}\,Q^{\frac{1}{2}}_{2m+2}}{Q_{2m+1}}
\end{gathered}\end{equation}
This gives equality \eqref{E1.20}
\end{proof}
The differential - difference hierarchies \eqref{E1.20},
\eqref{E1.21}, \eqref{E1.22} and \eqref{E1.23} are useful for
finding the special polynomials $Q_{n}(z)$ and $R_{n}(z)$ associated
with rational solutions of hierarchy \eqref{ku0.1}.

\section{Special polynomials associated with the
first member of hierarchy \eqref{ku0.1}}

The expressions $H_{1}[v]$ and $G_{1}[u]$ take the form
\begin{equation}\begin{gathered}
\label{ku0.10a}H_{1}[v]=v_{zz}+\,{4}\,v^{2},\,\,\quad\,\,G_{1}[u]=u_{zz}+\frac{1}{4}\,u^{2},
\end{gathered}\end{equation}

\begin{table}[h]%[t h]
    \center
    \caption{Polynomials $Q^{(1)}_n(z)$} \label{t:k4.1}
    \begin{tabular}{l} %{||c|c|c|c|c|c|c||}     %{||c|c|p{65mm}||}
        \hline
        $ Q^{(1)}_0=1 $, \\
        $ Q^{(1)}_1=z $, \\
        $ Q^{(1)}_2=z^4$, \\
        $ Q^{(1)}_3=z^8 $, \\
        $ Q^{(1)}_4=z^4\,(z^5-504)^2 $, \\
        $ Q^{(1)}_5 = z\, ( {z}^{20}-3276\,{z}^{15}+6604416\,{z}^{10} +3328625664 \,{z}^{5}-
        119830523904 ) $, \\
        $ Q^{(1)}_{{6}}= ({z}^{15}-6552\,{z}^{10}-13208832\,{z}^{5}-951035904)^{2} $, \\
        $ Q^{(1)}_{{7}}={z}^{40}-29952\,{z}^{35}+203793408\,{z}^{30}+3066139754496\,
        {z}^{25}+$ \\
        \qquad $ +5234197284126720\,{z}^{20}+36006491762989203456\,{z}^{15}- $ \\
        \qquad $ -3574462636834928197632\,{z}^{10}-7206116675859215246426112\,{z}^{5}+ $ \\
        \qquad $ + 129710100165465874435670016 $, \\
        $ Q^{(1)}_{{8}}={z}^{2}({z}^{25}-37440\,{z}^{20}-179262720\,{z}^{15}-
        4698117365760\,{z}^{10}- $ \\
        \qquad $ -1500270490124550144)^{2} $, \\
        $ Q^{(1)}_{{9}}={z}^{65}-142272\,{z}^{60}+5244715008\,{z}^{55}+18447301656576
        \,{z}^{50}+         $ \\
        \qquad $ +5585422603926896640\,{z}^{45}-22828544426446619148288\,{z}^{40}+$ \\
        \qquad $ +1478238865378129843895402496\,{z}^{35}+ $ \\
        \qquad $ +22449112629907483670818980888576\,{z}^{30}- $ \\
        \qquad $ -40520916747771106259841854742724608\,{z}^{25}+ $ \\
        \qquad $ +5403734062860426731139723747278192640\,{z}^{20}- $ \\
        \qquad $ -588158921353048128175262650355386536689664\,{z}^{15}- $ \\
        \qquad $ -38747522650600443047392623159427256373215232\,{z}^{10}+ $ \\
        \qquad $ +117172508495415739775315292434108023272602861568\,{z}^{5}+ $ \\
        \qquad $ +1406070101944988877303783509209296279271234338816$\\
        \hline
    \end{tabular}
\end{table}

\begin{table}[h]%[t h]
    \center
    \caption{Polynomials $R^{(1)}_n(z)$} \label{t:k4.2}
    \begin{tabular}{l}
        \hline
        $ R^{(1)}_0=1 $, \\
        $ R^{(1)}_1=z^2 $, \\
        $ R^{(1)}_2=z^5+36 $, \\
        $ R^{(1)}_3=(z^5-144)^2 $, \\
        $ R^{(1)}_{{4}}=z \left( {z}^{15}-1152\,{z}^{10}+1824768\,{z}^{5}+131383296\right)$,\\
        $ R^{(1)}_{{5}}={z}^{4} \left( {z}^{10}-3168\,{z}^{5}-3193344 \right) ^{2} $, \\
        $ R^{(1)}_{{6}}={z}^{8} \left({z}^{25}-15840\,{z}^{20}+63866880\,{z}^{15}+
        708155965440\,{z}^{10}+ \right. $ \\
        \qquad $ \left.+1922177762806726656 \right) $, \\
        $R^{(1)}_{{7}}={z}^{4} ({z}^{20}-22176\,{z}^{15}-95001984\,{z}^{10}-
        902898855936\,{z}^{5}+$ \\
        \qquad $ +303374015594496) ^{2} $, \\
        $ R^{(1)}_{{8}}=z\, ({z}^{55}-88704\,{z}^{50}+1900039680\,{z}^{45}+
        21067639971840\,{z}^{40}+$\\
        \qquad $ +1029196347904327680\,{z}^{35}-3888565614158008025088\,{z}^{30}+ $ \\
        \qquad $ +119982378242306659615506432\,{z}^{25}+ $ \\
        \qquad $ +745508187107335699834161070080\,{z}^{20}- $ \\
        \qquad $ -865913009382826100820052971356160\,{z}^{15}+ $ \\
        \qquad $ + 1723009449942845244977210857913057280\,{z}^{10}+ $ \\
        \qquad $ + 603307224662092676093915178711788814336\,{z}^{5}- $ \\
        \qquad $ - 14479373391890224226253964289082931544064) $, \\
        $ R^{(1)}_{{9}}=({z}^{35}-94248\,{z}^{30}-95001984\,{z}^{25}- 60569464919040\,{z}^{20}-
        $ \\
        \qquad $ -1037918350852669440\,{z}^{15}+ 4274562936151633821696\,{z}^{10}+ $ \\
        \qquad $ +6463139159461270338404352\,{z}^{5}+ 310230679654140976243408896)^{2} $\\
        \hline
    \end{tabular}
\end{table}

Substituting \eqref{ku0.10a} into hierarchies \eqref{ku0.1} and
\eqref{ku0.6} at $N=1$ we have the fourth - order equation
\cite{Hone01, Kudryashov07,
 Mugan01, Gromak01, Cosgrove01}
\begin{equation}\begin{gathered}
\label{ku0.14}w_{zzzz}+5\,w_{z}\,w_{zz}-5\,w^2\,w_{zz}-5\,w\,w_{z}^2+w^5-z\,w-\beta_1=0
\end{gathered}\end{equation}
Assuming $Q^{(1)}_{0}(z)=R^{(1)}_{0}=1$, $Q^{(1)}_{1}=z$,
$R^{(1)}_{1}=z^2$ and using the differential - difference
hierarchies \eqref{E1.20}, \eqref{E1.21}, \eqref{E1.22} and
\eqref{E1.23} we find the special polynomials $Q^{(1)}_{n}(z)$ and
$R^{(1)}_{n}(z)$ at $n\geq 2$. The first nine special polynomials
$Q^{(1)}_{n}(z)$ and $R^{(1)}_{n}(z)$ are given in Tables
\eqref{t:k4.1} and \eqref{t:k4.2}

\begin{figure}[h]
 \centerline{\epsfig{file=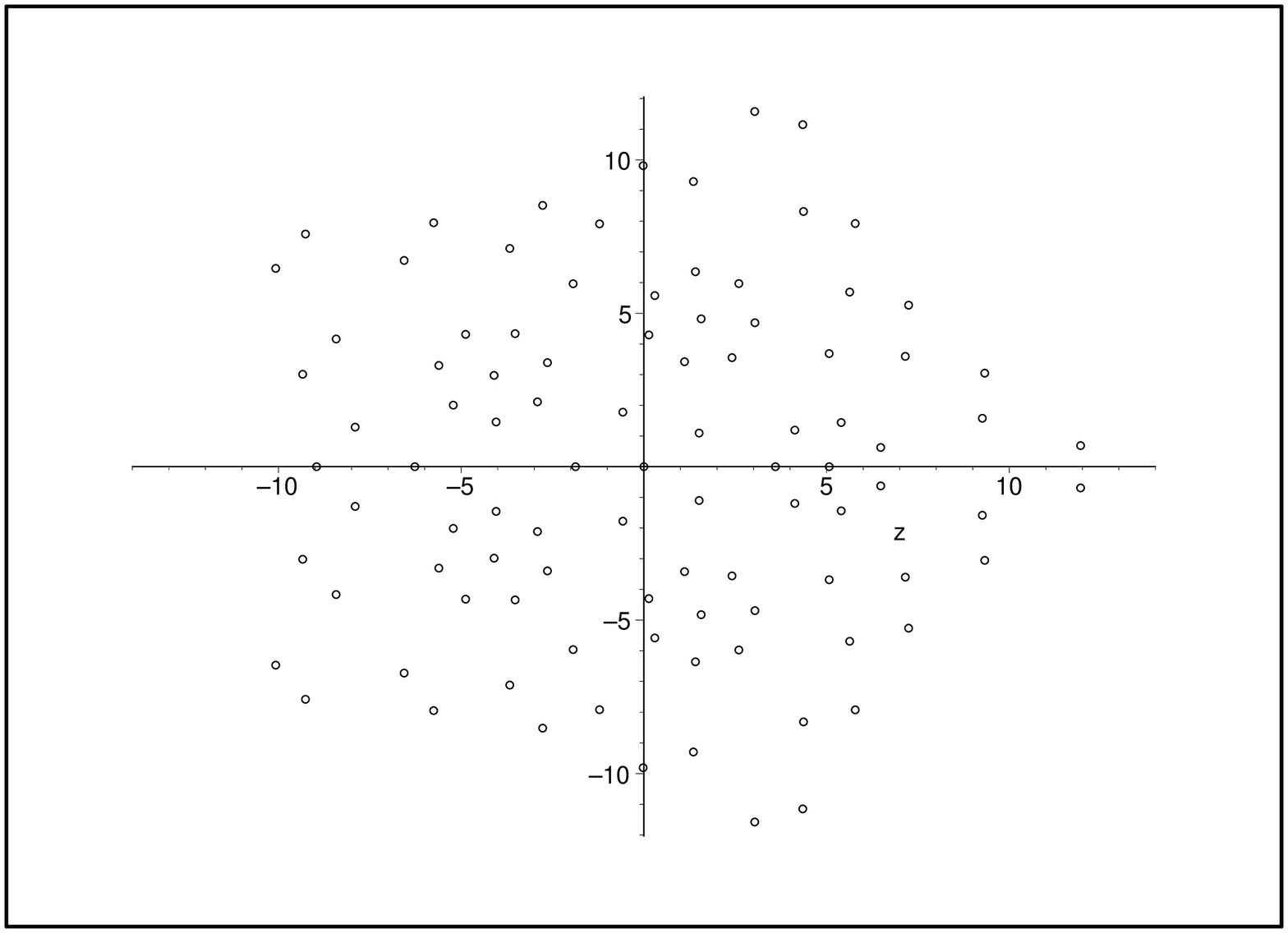,angle=270,width=120mm}}
 \caption{Roots of polynomial $Q^{(1)}_{11}(z)=0$}\label{fig3:z_post}
\end{figure}

\begin{figure}[h]
 \centerline{\epsfig{file=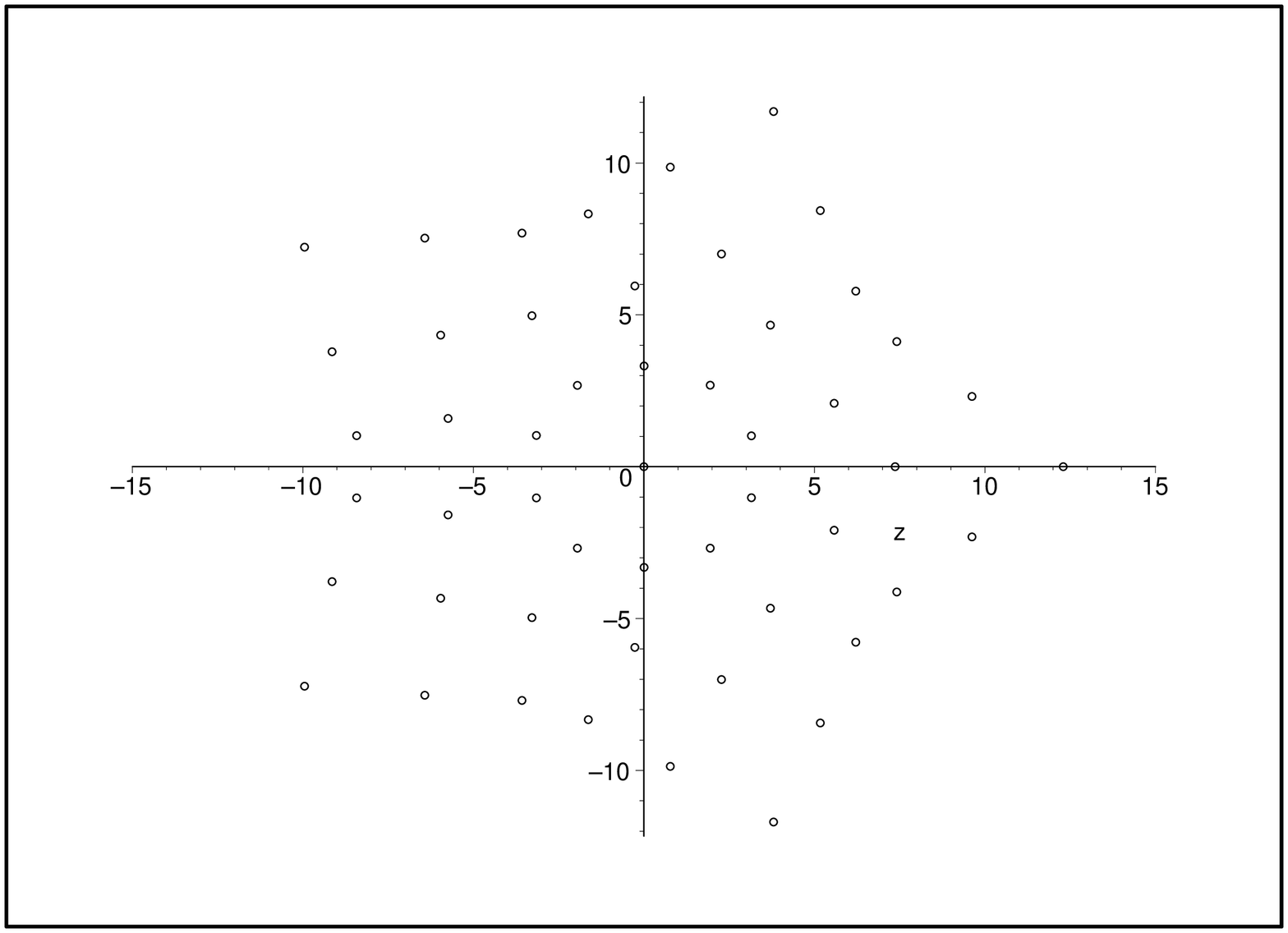,angle=270,width=120mm}}
 \caption{Roots of polynomial $R^{(1)}_{11}(z)=0$}\label{fig7:z_post}
\end{figure}

Recently Clarkson and Mansfield \cite{Clarkson01} studied the
structure of the roots of the Yablonskii - Vorob'ev polynomials and
showed that they have a highly regular pattern which is very
symmetric and structured. In Figure 1 and Figure 2 the locations of
the roots for the polynomials $Q^{(1)}_{11}(z)=0$ and
$R^{(1)}_{11}(z)=0$ are plotted. From these plots we observe that
the roots of the polynomials form approximately regular pentagons.
In the case of the roots for the polynomials $Q^{(1)}_{11}(z)=0$ we
have the cutting angle but for the polynomials $R^{(1)}_{11}(z)=0$
we get the acute angle. From plots we can see that the roots of the
polynomials form a highly symmetric structures.

Using the special polynomials $Q^{(1)}_{n}(z)$ and $R^{(1)}_{n}(z)$
we find the rational solutions of equation \eqref{ku0.14} by means
of formulae \eqref{ku0.5a}.

\begin{equation}\begin{gathered}
\label{3.1}w{{(z;-1)}}=\frac{1}{z},\,\,\quad\,w{{(z;3)}}=-\frac{3}{z},\,\,\quad\,
w{{(z;-4)}}=\frac{4}{z},\\
\\
w{{(z;6)}}=-\,{\frac {6\,({z}^{5}+336)}{z \left( {z}^{5}-504 \right)
}},
\end{gathered}\end{equation}

\begin{equation}\begin{gathered}
\label{3.5}w{{(z;2)}}=-\frac{2}{z},\,\,\quad\,w{{(z;-3)}}=\,{\frac
{3({z}^{5}-24)}{z \left( {z}^{5}+36 \right) }},\,\quad
w(z;5)=-\frac{5z^4(z^5+216)}
{(z^5+36)(z^5-144)},\\
\\
w{{(z;-6)}}={\frac
{6({z}^{20}-576\,{z}^{15}-912384\,{z}^{10}-459841536\,{z}^{5}-3153199104)}{z
\left( {z}^{5}-144 \right)  \left( {
z}^{15}-1152\,{z}^{10}+1824768\,{z}^{5}+131383296 \right) }}
\end{gathered}\end{equation}

We have obtained solutions \eqref{3.1}  using the special
polynomials $Q^{(1)}_{n}(z)$ from table (4.1) and we find solutions
\eqref{3.5} taking into account the special polynomials
$R^{(1)}_{n}(z)$ from table (5.2).

\section{Special polynomials associated with the
second member of hierarchy \eqref{ku0.1}}

From \eqref{ku0.2} and \eqref{ku0.7} we have
\begin{equation}\begin{gathered}
\label{ku0.10}H_{2}[v]=v_{zzzz}+12\,v\,v_{zz}+6\,v^{2}_{z}
+\frac{32}{3}\,v^3,
\end{gathered}\end{equation}
\begin{equation}\begin{gathered}
\label{ku0.11}G_2[u]=u_{zzzz}+\frac{3}{2}\,u\,u_{zz}+\frac{3}{4}\,u^{2}_{z}+
\frac{1}{6}\,u^3,
\end{gathered}\end{equation}
Substituting \eqref{ku0.10} and \eqref{ku0.11} into \eqref{ku0.1}
and \eqref{ku0.6} at $n=2$ yields the sixth - order equation
\begin{equation}\begin{gathered}
\label{ku0.15}w_{zzzzzz}+7\,w_{z}\,w_{zzzz}-20\,w_{z}^2\,w_{zz}-21\,w\,w_{zz}^2+
14\,w_{zz}\,w_{zzz}-
\\
-7\,w^3\,w_{zzzz}-\frac{28}{3}\,w\,w_{z}^2+14\,w^4\,w_{zz}+28\,w^3\,w_{z}^2-
28\,w\,w_{z}\,w_{zzz}-
\\
-14\,w^2\,w_z\,w_{zz}-\frac{4}{3}\,w^7-z\,w-\beta_2=0
\end{gathered}\end{equation}

\begin{table}[h]%[t h]
    \center
    \caption{Polynomials $Q^{(2)}_n(z)$} \label{t:k5.1}
    \begin{tabular}{l} %{||c|c|c|c|c|c|c||}     %{||c|c|p{65mm}||}
        \hline
        $ Q^{(2)}_0=1 $, \\
        $ Q^{(2)}_1=z $, \\
        $ Q^{(2)}_2=z^4$, \\
        $ Q^{(2)}_3=z \left( {z}^{7}-1728 \right) $, \\
        $ Q^{(2)}_4=\left( {z}^{7}+4320 \right) ^{2}$, \\
        $ Q^{(2)}_5 = {z}^{21}+43200\,{z}^{14}-2426112000\,{z}^{7}+2620200960000 $, \\
        $ Q^{(2)}_{{6}}= {z}^{2} \left( {z}^{14}+280800\,{z}^{7}+4852224000 \right) ^{2} $, \\
        $ Q^{(2)}_{{7}}={z}^{5}( {z}^{35}+1684800\,{z}^{28}+1610938368000\,{z}^{21}-
        632202087628800000\,{z}^{14}- $ \\
        \qquad $-46523097626443776000000\,{z}^{7}+803919126984948449280000000 ),$ \\
        $ Q^{(2)}_{{8}}={z}^{10} \left( {z}^{21}+3369600\,{z}^{14}-2396998656000\,{z}^{7}+
        455621504532480000 \right) ^{2} $, \\
        $ Q^{(2)}_{{9}}={z}^{16}( {z}^{49}+15724800\,{z}^{42}+67115962368000\,{z}^{35}-
        342130329767116800000\,{z}^{28}-$ \\
        \qquad $-136126583654974488576000000\,{z}^{21}-
        1009122699824364480752517120000000\,{z}^{14}+$ \\
         \qquad $+121829319173072454489913623891148800000000000)$\\
         \hline
    \end{tabular}
\end{table}

\begin{table}[h]%[t h]
    \center
    \caption{Polynomials $R^{(2)}_n(z)$} \label{t:k5.2}
    \begin{tabular}{l}
        \hline
        $ R^{(2)}_0=1 $, \\
        $ R^{(2)}_1=z^2 $, \\
        $ R^{(2)}_2=z^5$, \\
        $ R^{(2)}_3=z^{10}$, \\
        $ R^{(2)}_{{4}}=z^{16}$, \\
        $ R^{(2)}_{{5}}={z}^{10} \left( {z}^{7}+95040 \right) ^{2} $, \\
        $ R^{(2)}_{{6}}={z}^{8} \left({z}^{25}-15840\,{z}^{20}+63866880\,{z}^{15}+
        708155965440\,{z}^{10}+ \right. $ \\
        \qquad $ \left.+1922177762806726656 \right) $, \\
        $ R^{(2)}_{{7}}={z}^{2} \left( {z}^{21}+1615680\,{z}^{14}-383885568000\,{z}^{7}+
        3316771307520000 \right) ^{2} $,\\
        $ R^{(2)}_{{8}}={z}^{56}+8078400\,{z}^{49}+21497591808000\,{z}^{42}-
        39469578559488000000\,{z}^{35}+$\\
        \qquad $+3188510434349678592000000\,{z}^{28}-
        9839758205158093309870080000000\,{z}^{21}+ $\\
        \qquad $+301096601077837655282024448000000000\,{z}^{14}+ $ \\
        \qquad $+8176063133267911645143886725120000000000\,{z}^{7}+ $ \\
        \qquad $+4415074091964672288377698831564800000000000$,\\
        $ R^{(2)}_{{9}}=( {z}^{35}+12117600\,{z}^{28}-35317472256000\,{z}^{21}+
        9726431859302400000\,{z}^{14}+ $ \\
        \qquad $+8011467393870200832000000\,{z}^{7}-17304769570759633797120000000) ^{2},$ \\
        \hline
    \end{tabular}
\end{table}

\begin{figure}[h]
 \centerline{\epsfig{file=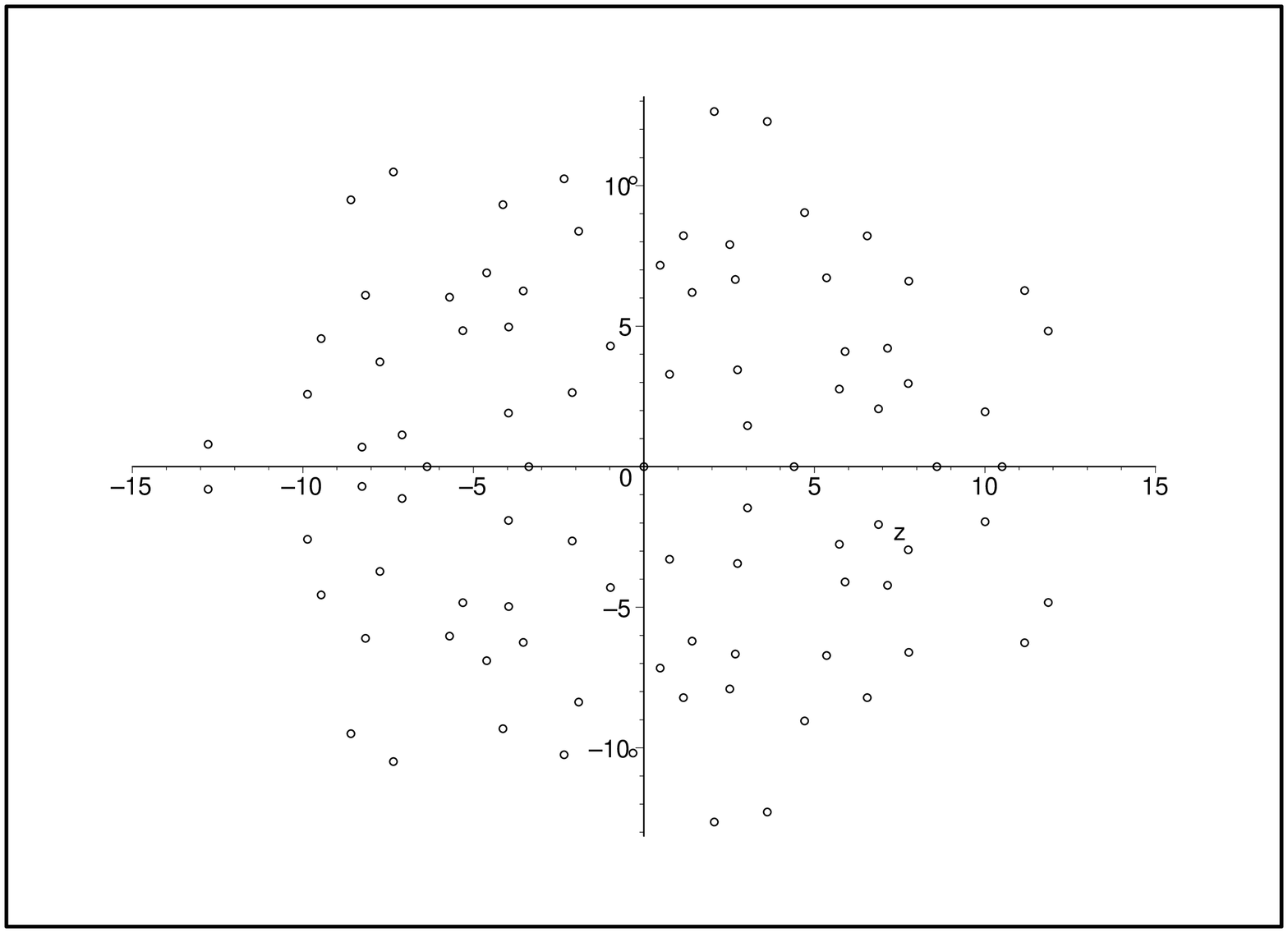,angle=270,width=120mm}}
 \caption{Roots of polynomial $Q^{(2)}_{11}(z)=0$}\label{fig9:z_post}
\end{figure}

\begin{figure}[h]
 \centerline{\epsfig{file=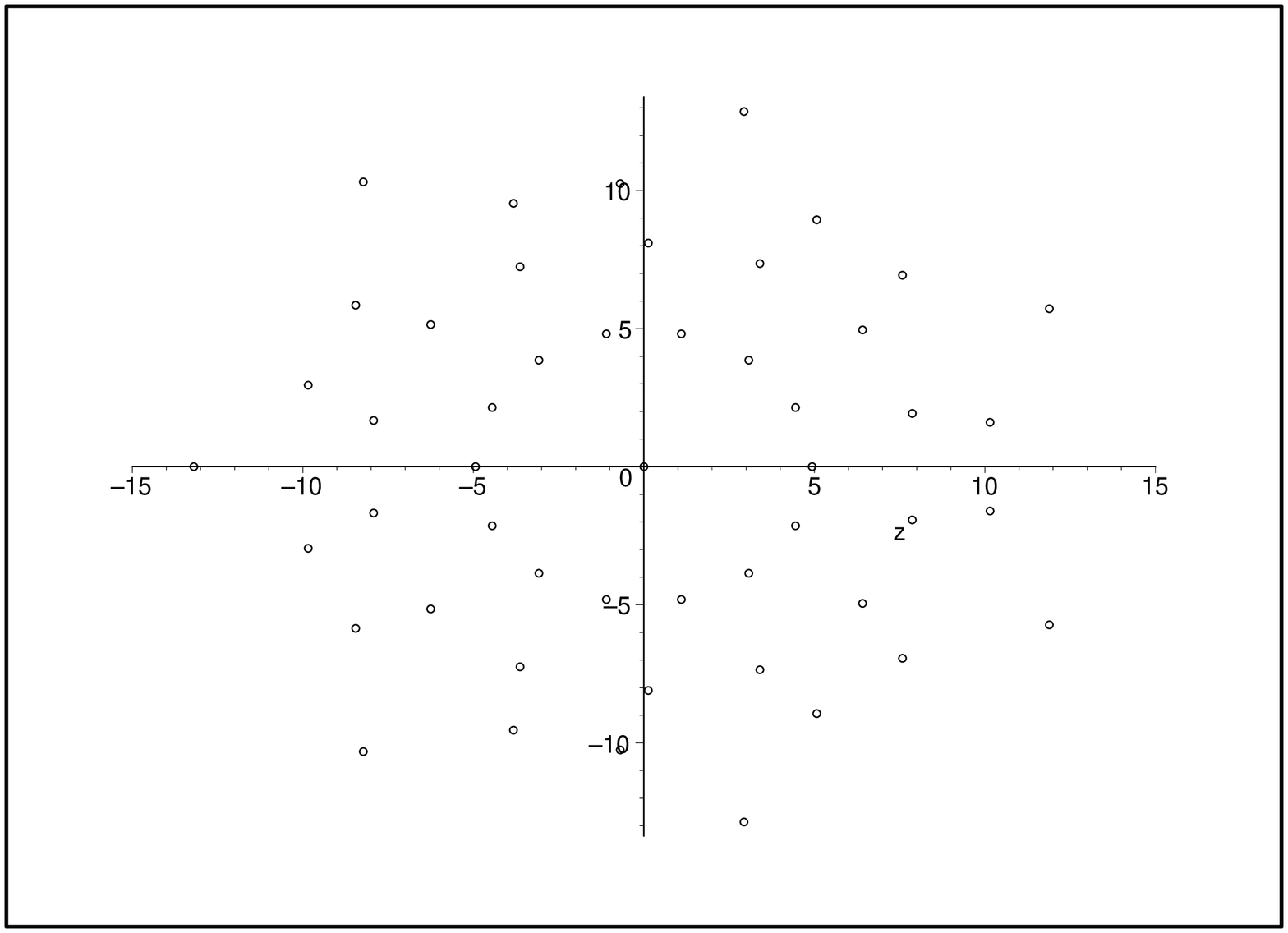,angle=270,width=120mm}}
 \caption{Roots of polynomial $R^{(2)}_{11}(z)=0$}\label{fig11:z_post}
\end{figure}

Taking  $Q^{(2)}_{0}(z)=R^{(2)}_{0}=1$, $Q^{(2)}_{1}=z$,
$R^{(2)}_{1}=z^2$ and using the differential - difference
hierarchies \eqref{E1.20}, \eqref{E1.21}, \eqref{E1.22} and
\eqref{E1.23} we get the special polynomials $Q^{(2)}_{n}(z)$ and
$R^{(2)}_{n}(z)$ at $n\geq 2$. The first  nine special polynomials
are given in Tables \eqref{t:k5.1} and \eqref{t:k5.2}.

In Figure 3 and Figure 4 the locations of the roots for the
polynomials $Q^{(2)}_{11}(z)=0$ and $R^{(2)}_{11}(z)=0$ are plotted.
From these plots we observe that the roots of the polynomials form
approximately regular heptagon.

Using the special polynomials $Q^{(2)}_{n}(z)$ and $R^{(2)}_{n}(z)$
we get the rational solutions of equation \eqref{ku0.15} in the form

\begin{equation}\begin{gathered}
\label{6.1}w{{(z;-1)}}=\frac{1}{z},\,\,\quad\,\,w{{(z;3)}}=-\frac{3}{z},\,\,\quad\,\,
w{{(z;-4)}}=\,{\frac {4\,({z}^{7}+1296)}{z \left( {z}^{7}-1728
\right) }},\\
\\
w{{(z;6)}}=-\,{\frac {6\,({z}^{14}-9504\,{z}^{7}+1244160)}{ \left(
{z}^{7}-1728 \right) z \left( {z}^{7}+4320 \right) }},
\end{gathered}\end{equation}

\begin{equation*}\begin{gathered}
\label{6.3}w{{(z;-7)}}={\frac { 7\left(
{z}^{21}+12960{z}^{14}+2799360000{z}^{7}- 15721205760000 \right)
{z}^{6}}{ \left( {z}^{7}+4320 \right) \left( {
z}^{21}+43200{z}^{14}-2426112000{z}^{7}+2620200960000 \right)}},
\end{gathered}\end{equation*}

\begin{equation}\begin{gathered}
\label{6.4}\quad\,\,w{{(z;2)}}=-\frac{2}{z},\,\,\,\quad\,\,
w{{(z;-3)}}=\frac{3}{z},\,\,\quad\,\,w{{(z;5)}}=-\frac{5}{z},\\
\\
w{{(z;-6)}}=\frac{6}{z},\,\,\quad\,\,w{{(z;8)}}=-{\frac
{8\,({z}^{7}-71280)}{z \left( {z}^{7}+95040 \right) }}
\end{gathered}\end{equation}

We have found solutions \eqref{6.1} using the special polynomials
$Q^{(2)}_{n}(z)$ from table (5.1) and we obtained solutions
\eqref{6.4} taking into account the special polynomials
$R^{(2)}_{n}(z)$ from table (5.2).

\section{Special polynomials associated with the
third member of hierarchy \eqref{ku0.1}}

Using  \eqref{ku0.3}  and \eqref{ku0.8} we have from \eqref{ku0.2}
and \eqref{ku0.7}

\begin{equation}\begin{gathered}
\label{ku0.12}H_{3}[v]=v_{zzzzzzzz}+20\,v\,v_{{{zzzzzz}}}+60\,v_{{z}}v_{{{
zzzzz }}}+134\,v_{{{zz}}}\,v_{{{zzzz}}}+
\\
+136\,{v}^{2}v_{{{zzzz}}}+
84\,{v^{2}_{{{zzz}}}}+544\,v\,v_{{z}}v_{{{zzz}}}+408\,v{v^{2}_{{{
zz}}}}+396\,{v^{2}_{{z}}}v_{{{zz}}}+
\\
+{\frac {1120}{3}}\,{v}
^{3}v_{{{zz}}}+560\,{v}^{2}{v^{2}_{{z}}}+{\frac {256}{3}}\,{v}^{5},
\end{gathered}\end{equation}
\begin{equation}\begin{gathered}
\label{ku0.13}G_3[u]=u_{zzzzzzzz}+\frac{7}{2}\,u\,u_{{{zzzzzz}}}+\frac{21}{2}\,u_{{z}}
u_{{{zzzzz}}}+{\frac{37}{2}}\,u_{{{zz}}}u_{ {{ zzzz}}}+
\\
+4\,{u}^{2}\,u_{{{zzzz}}}++{\frac {39}{4}}\,{u^{2}_{{{
zzz}}}}16\,u\,u_{{z}}\,u_{{{zzz}}}+12\,u{u^{2}_{{{zz}}}}+{\frac
{33}{2}}\,{u^{2}_{{z}}}u_{{{zz}}}+
\\
+{\frac {25}{ 12}}\,{u}^{3}\,u_{{{zz}}}+{\frac {25}{8}}
\,{u}^{2}{u^{2}_{{z}}}+\frac{1}{12}\,{u}^{5}
\end{gathered}\end{equation}

\begin{table}[h]%[t h]
    \center
    \caption{Polynomials $Q^{(3)}_n(z)$} \label{t:k6.1}
    \begin{tabular}{l} %{||c|c|c|c|c|c|c||} %{||c|c|p{65mm}||}
        \hline
        $ Q^{(3)}_0=1 $, \\
        $ Q^{(3)}_1=z $, \\
        $ Q^{(3)}_2=z^4$, \\
        $ Q^{(3)}_3=z^8 $, \\
        $ Q^{(3)}_4= z^{14}$, \\
        $ Q^{(3)}_5 = {z}^{21}$, \\
        $ Q^{(3)}_{{6}}={z}^{30}$, \\
        $ Q^{(3)}_{{7}}={z}^{40}$, \\
        $ Q^{(3)}_{{8}}= {z}^{30} \left( {z}^{11}+21511526400 \right) ^{2}$, \\
        $ Q^{(3)}_{{9}}={z}^{21}({z}^{44}+153653760000\,{z}^{33}+
        23137288402894848000000\,{z}^{22}-$ \\
        \qquad$-124429597575821589793996800000000\,
        {z}^{11}-$\\
        \qquad $-281754797178280224267740286812160000000000)$\\
                \hline
    \end{tabular}
\end{table}

\begin{table}[h]%[t h]
    \center
    \caption{Polynomials $R^{(3)}_n(z)$} \label{t:k6.2}
    \begin{tabular}{l}
        \hline
        $ R^{(3)}_0=1 $, \\
        $ R^{(3)}_1=z^2 $, \\
        $ R^{(3)}_2=z^5$, \\
        $ R^{(3)}_3=z^{10}$, \\
        $ R^{(3)}_{{4}}={z}^{16}$, \\
        $ R^{(3)}_{{5}}= {z}^{2} \left( {z}^{11}-43545600 \right) ^{2}$,\\
        $ R^{(3)}_{{6}}= {z}^{33}-609638400\,{z}^{22}-90260037697536000\,{z}^{11}-
        491303437195227955200000,$ \\
        $ R^{(3)}_{{7}}=\left( {z}^{22}+5181926400\,{z}^{11}-112825047121920000 \right) ^
        {2},
        $ \\
        $ R^{(3)}_{{8}}=z( {z}^{55}+41455411200\,{z}^{44}+4280582287805644800000\,{z}^{33}-
        $\\
        \qquad $-4408269220577902350827520000000\,{z}^{22}-$ \\
        \qquad $-1045218527199591396510535778304000000000\,{z}^{11}- $ \\
        \qquad $-7585777983003754519314864464619110400000000000),$\\
        $ R^{(3)}_{{9}}={z}^{4}({z}^{33}+207277056000\,{z}^{22}-
        4632032309590425600000 \,{z}^{11}+$ \\
        \qquad $+437026689971085146849280000000) ^{2}$\\
              \hline
    \end{tabular}
\end{table}

Substituting \eqref{ku0.12} and \eqref{ku0.13} into equations
\eqref{ku0.1} and \eqref{ku0.6} we find the tenth - order equation
at $N=3$ (This equation is presented in the appendix A).

Assuming $Q^{(3)}_{0}(z)=R^{(3)_{0}}=1$, $Q^{(3)}_{1}=z$,
$R^{(3)}_{1}=z^2$ and using the differential - difference hierarchy
\eqref{E1.20}, \eqref{E1.21}, \eqref{E1.22} and \eqref{E1.23} we get
the special polynomials $Q^{(3)}_{n}(z)$ and $R^{(3)}_{n}(z)$ at
$n\geq 2$. The first nine special polynomials are given in Tables
\eqref{t:k6.1} and \eqref{t:k6.2}.

In Figure 5 and Figure 6 the locations of the roots for the
polynomials $Q^{(3)}_{11}(z)=0$ and $R^{(3)}_{11}(z)=0$ are plotted.
From these plots we observe that the roots of the polynomials form
approximately regular polygons with eleven angles. Studying other
plots we observe the symmetric plots for other roots of the
polynomials $Q^{(3)}_{n}(z)=0$ and $R^{(3)}_{n}(z)=0$.

Using the special polynomials $Q^{(3)}_{n}(z)$ and $R^{(3)}_{n}(z)$
we get the rational solutions $w(z;\beta)$ of the tenth - order
equation in the form

\begin{figure}[h]
 \centerline{\epsfig{file=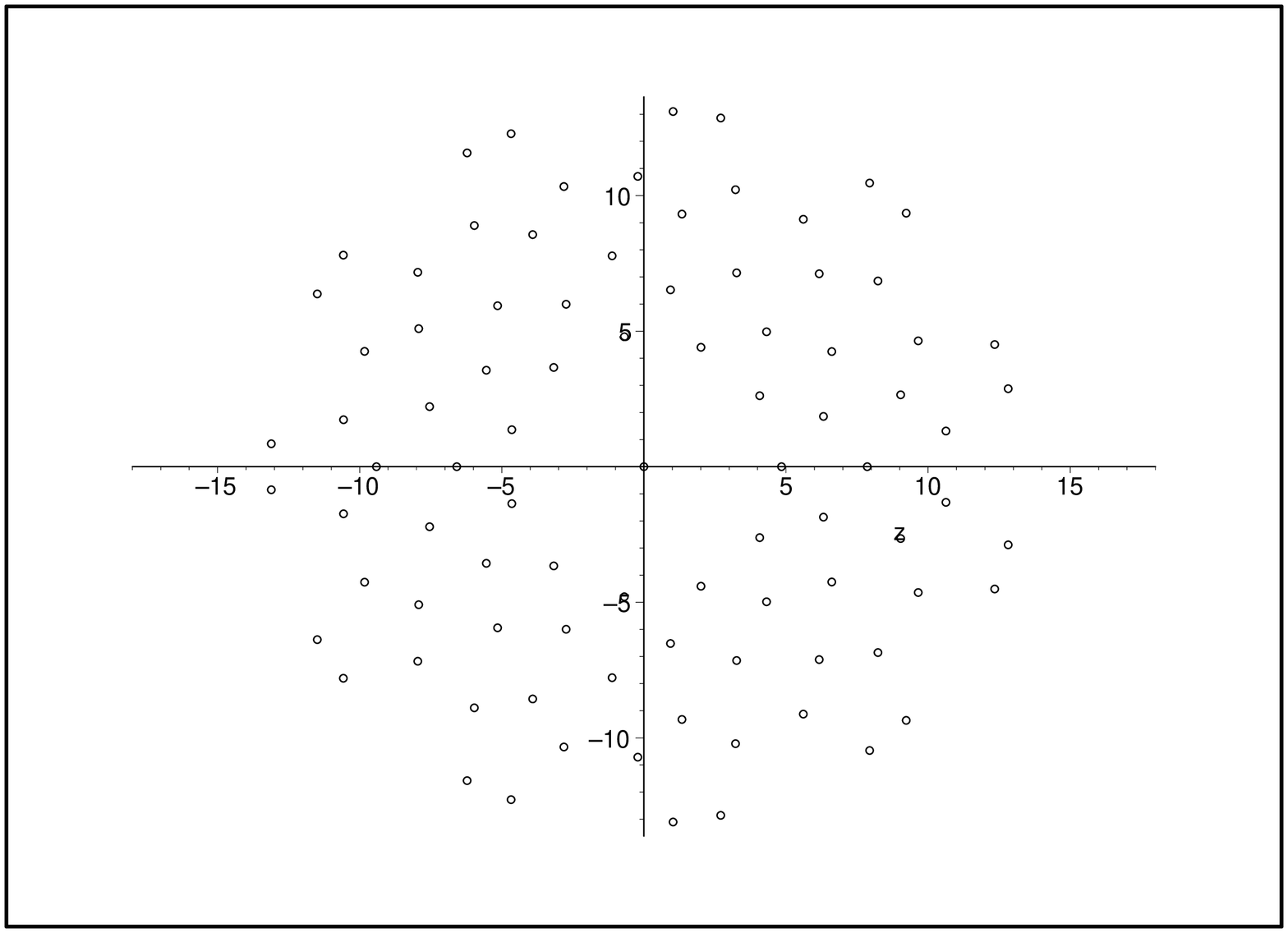,angle=270,width=120mm}}
 \caption{Roots of polynomial $Q^{(3)}_{11}(z)$}\label{fig13:z_post}
\end{figure}

\begin{figure}[h]
 \centerline{\epsfig{file=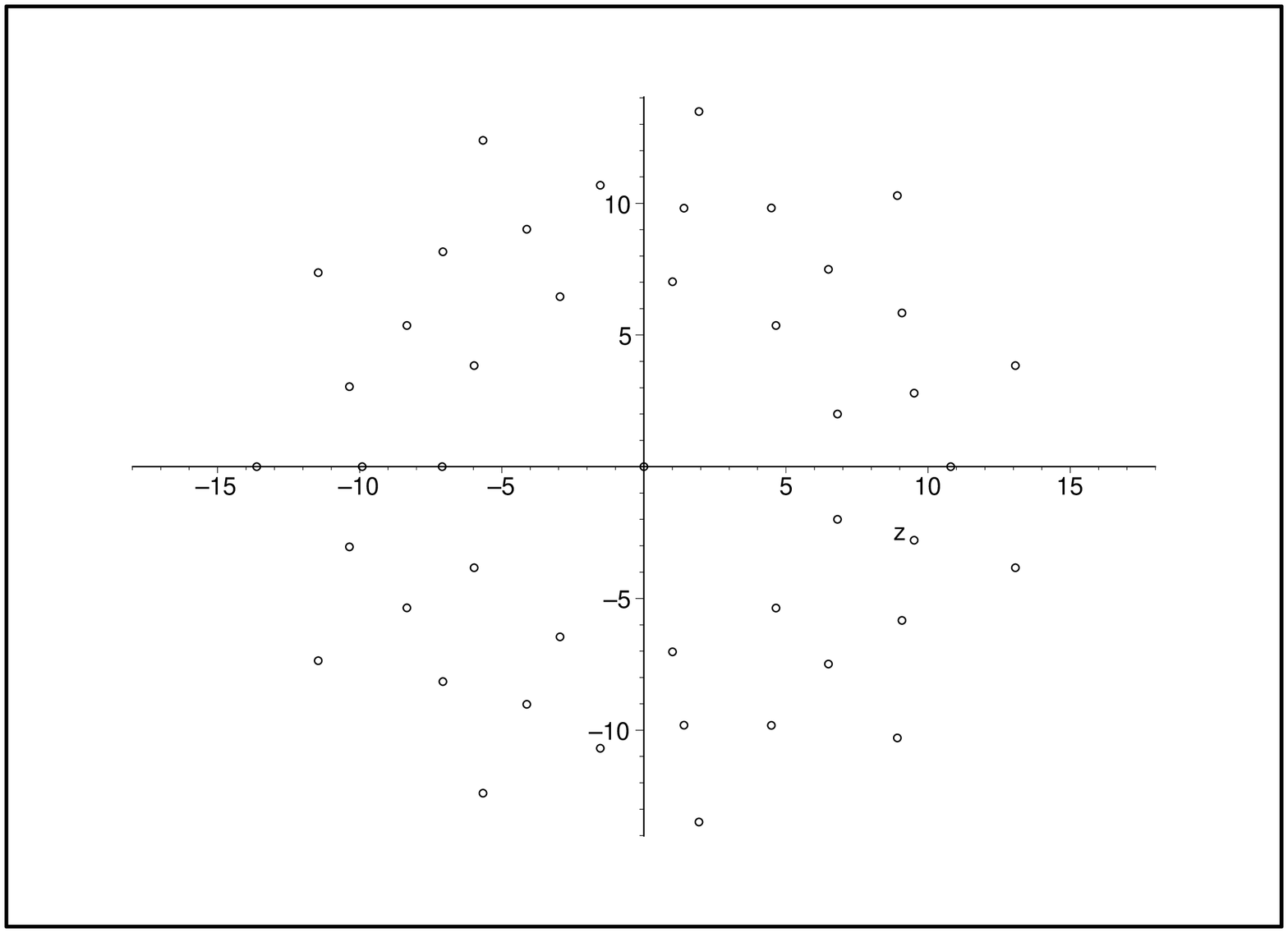,angle=270,width=120mm}}
 \caption{Roots of polynomial $R^{(3)}_{11}(z)$}\label{fig15:z_post}
\end{figure}

\begin{equation}\begin{gathered}
\label{8.1}w{{(z;-1)}}=\frac{1}{z},\,\,\quad\,\,w{{(z;3)}}=-\frac{3}{z},\,\,\quad\,\,
w{{(z;-4)}}=\frac{4}{z},\,\,\quad\,\,w{{(z;6)}}=-\frac{6}{z}\\
\\
w{{(z;-7)}}=\frac{7}{z},\,\,\quad\,\,w{{(z;9)}}=-
\frac{9}{z},\,\,\quad\,\,w{{(z;-10)}}=\frac{10}{z},\\
\\
w{{(z;12)}}=-{\frac {12\,({z}^{11}-17926272000)}{z \left(
{z}^{11}+21511526400\right) }},
\end{gathered}\end{equation}

\begin{equation}\begin{gathered}
\label{8.3}w{{(z;2)}}=-\frac{2}{z},\,\,\quad\,\,w{{(z;-3)}}=\frac{3}{z},\,\,\quad\,\,
w{{(z;5)}}= -\frac{5}{z},\\
\\
w{{(z;-6)}}=\,{\frac {6\,({z}^{11}-20736000)}{z \left(
{z}^{11}+24883200 \right) }},\\
\\
w{{(z;8)}}=-{\frac
{8\,({z}^{22}+135302400\,{z}^{11}+406332702720000)}{z \left( {z}
^{11}+24883200 \right)  \left( {z}^{11}-43545600 \right) }}
\end{gathered}\end{equation}

We have got solutions \eqref{8.1} using the special polynomials
$Q^{(3)}_{n}(z)$ from table (6.1) and we found solutions \eqref{8.3}
taking into account the special polynomials $R^{(3)}_{n}(z)$ from
table (6.2).

\section {Conclusion}

We have introduced the special polynomials associated with the
rational solutions of hierarchy \eqref{ku0.1}. This hierarchy arises
if we look for the special solutions taking into account the scaling
reductions for the Fordy - Gibbons equation, the Caudrey - Dodd -
Gibbon hierarchy and the Kaup - Kupershmidt hierarchy. We have found
some recursion relations for the special polynomials of the
hierarchy studied. Using these relations we have obtained the
differential - difference hierarchies for finding the special
polynomials of the hierarchy considered. These recursion formulae
were used for finding the specials polynomials at the known two
previous special polynomials. The special polynomials of the first,
the second and the third members of the hierarchy were presented.
These special polynomials are new and similar to the Yablonskii -
Vorob'ev polynomials \cite{Yablonskii01, Vorob'ev01, Clarkson02,
Clarkson03}, the Okamoto polynomials \cite{Clarkson04, Okamoto01}
and the Umemura polynomials \cite{Umemura01} associated with the
rational solutions to the Painleve equations. Using the special
polynomials found the rational solutions for the first, the second
and the third members of the $K_2$ hierarchy were given.

We hope that the recursion relations \eqref{E1.1},  \eqref{E1.2},
\eqref{E1.3}, \eqref{E1.4}, \eqref{E1.10}, \eqref{E1.11},
\eqref{E1.12}, \eqref{E1.13} and the differentional - difference
hierarchies \eqref{E1.20}, \eqref{E1.21}, \eqref{E1.22} and
\eqref{E1.23} will be useful in proving various properties  for the
special polynomials associated with hierarchy \eqref{ku0.1}.

\section {Acknowledgments}

This work was supported by the International Science and Technology
Center under Project B 1213.

\appendix

\section {The third member of hierarchy \eqref{ku0.1}  }

\begin{equation}\begin{gathered}
\label{A.1}w_{{{10z}}}+11\,w_{{z}}w_{{{ 8z}}}-11\,{w}^{2}w_{{
{8z}}}+44\,w_{{{2z}}}w_{{{ 7z}}}-88\,w\,w_{{z}}w_{ {{7z}}}+
\\
\\
+110\,w_{{{3z}}}w_{{{
6z}}}-242\,ww_{{{2z}}}w_{{{6z}}}-66\,{w}^{2}w_{{z}}w_{{{6z}}}-198\,{w_{{z}}}^{2}w_{{{6z}}}+
\\
\\
+44\,{w}^{4}w_{{{
6z}}}+176\,w_{{{4z}}}w_{{{5z}}}-418\,ww_{{{3z}}}w_{{{5z}}}-990\,w_{{z}}w_{{{2z}}}w_{{{5z}}}-
\\
\\
-198\,{w}^{2}w_{{{2z}}}w_{{{ 5z}}}
-396\,w{w_{{z}}}^{2}w_{{{5z}}}+528\,{w}^{3}w_{{z}}w_{{{5z}}}-253\,w{w_{{{4z}}}}^{2}-
\\
\\
-330\,{w}^{2}w_{{{3z}}}w_{{{4z}}}-1518\,w_{{z}}w_{{{
3z}}}w_{{{4z}}}-1166\,{w_{{{2z}}}}^{2}w_{{{4z}}}+1100\,{w}^{3}w_{{{2z}}}w_{{{
4z}}}-
\\
\\
-1540\,ww_{{z}}w_{{{2z} }}w_{{{4z}}}-{\frac
{2090}{3}}\,{w_{{z}}}^{3}w_{{{4z}}}+
2310\,{w}^{2}{w_{{z}}}^{2}w_{{{4z}}}+110\,{w}^{4}w_{{z}}w_{{{4z}}}-
\\
\\
-{\frac
{242}{3}}\,{w}^{6}w_{{{4z}}}-1540\,w_{{{2z}}}{w_{{{3z}}}}^{2}-880\,ww_{{z}}{w_{{{
3z}}}}^{2}+660\,{w} ^{3}{w_{{{3z}}}}^{2}-
\\
\\
-1100\,w{w_{{{
2z}}}}^{2}w_{{{3z}}}-3300\,{w_{{z}}}^{2}w_{{{2z}}}w_{{{3z}}}+220\,{w}^{4}w_{{{
2z}}}w_{{{3z}}}+880\,{w}^{3}{w_{{z}}}^{2}w_{{{ 3z}}}+
\\
\\
+7920\,{w}^{2}w_{{z}}w_{{{2z}}}w_{{{3z}}}+3960\,w{w_{{z}}}^{3}w_{{{3z}}}-
968\,{w}^{5}w_{{z}}w_{{{3z}}}-1430\,w_{{z}}{w_{{{2z}}} }^{3}+
\\
\\
+1870\,{w}^{2}{w_{{{2z}}}}^{3}+1320\,{w}^{3}w_{{z}}{w_{{{
2z}}}}^{2}+8250\,w{w_{{z}}}^{2}{w_{{{2z}}}}^{2}-726\,{w}^{5}{w_{{{
2z}}}}^{2}-
\\
\\
-4620\,{w}^{4}w_{{{2z}}}{w_{{z}}}^{2}+{\frac {9460}{
3}}\,{w_{{z}}}^{3}{w}^{2}w_{{{2z}}}-{\frac {220}{3}}\,{w}^{6}w_{{{
2z}}}w_{{z}}+{\frac {7040}{3}}\,w_{{{2z}}}{w_{{z}}}^{4}+
\\
\\
+{ \frac {220}{3}}\,{w}^{8}w_{{{2z}}}+{\frac
{2200}{3}}\,w{w_{{z}}}^{ 5}-{\frac
{440}{3}}\,{w}^{5}{w_{{z}}}^{3}+{\frac {880}{3}}\,{w}^{7}{w_
{{z}}}^{2}-
\\
\\
-2200\,{w}^{3}{w_{{z}}}^{4}-\frac83\,{w}^{11}-z\,w-\beta_{{3}}
=0,\,\qquad\,\,w_{nz}= \frac{d^{n}w}{dz^n}
\end{gathered}\end{equation}

\end{document}